\documentclass[graphics,twocolumn,floatfix,mathbbm,pra,superscriptaddress,aps]{revtex4}
\usepackage{amsmath,amssymb,graphicx,color}
\usepackage{float}

	\newcommand{\Prs}{$\text{Pr}^{3+}\text{:Y}_2\text{SiO}_5$ }
	
	\newcommand{\3}{$^{3+}$}

	\newcommand{\hac}{$g^{(2)}_{i:s,s}$\,}
	\newcommand{\gcc}{$g^{(2)}_{s,i}$\,}
	
	\newcommand{\gc}{$g^{(2)}_{s,i}$}
	
		\newcommand{\numspmodes}{15\,}
		\newcommand{\ii}{\emph{i} }
		\newcommand{\iin}{\emph{i}}		
		\newcommand{\s}{\emph{s} }
		\newcommand{\sn}{\emph{s}}

	\definecolor{darkgreen}{RGB}{0,170,50}

\begin{document}

\title{Quantum Storage of Frequency-Multiplexed Heralded Single Photons}

\author{Alessandro Seri}
\altaffiliation{These authors contributed equally to this paper}
\affiliation{ICFO-Institut de Ciencies Fotoniques, The Barcelona Institute of Technology, Mediterranean Technology Park, 08860 Castelldefels (Barcelona), Spain}
\email{alessandro.seri@icfo.es}
\author{Dario Lago}
\altaffiliation{These authors contributed equally to this paper}
\affiliation{ICFO-Institut de Ciencies Fotoniques, The Barcelona Institute of Technology, Mediterranean Technology Park, 08860 Castelldefels (Barcelona), Spain}
\author{Andreas Lenhard}
\affiliation{ICFO-Institut de Ciencies Fotoniques, The Barcelona Institute of Technology, Mediterranean Technology Park, 08860 Castelldefels (Barcelona), Spain}
\author{Giacomo Corrielli}
\affiliation{Istituto di Fotonica e Nanotecnologie (IFN) - CNR and Dipartimento di Fisica - Politecnico di Milano, P.zza Leonardo da Vinci 32, 20133 Milano, Italy}
\author{Roberto Osellame}
\affiliation{Istituto di Fotonica e Nanotecnologie (IFN) - CNR and Dipartimento di Fisica - Politecnico di Milano, P.zza Leonardo da Vinci 32, 20133 Milano, Italy}
\author{Margherita Mazzera}
\affiliation{ICFO-Institut de Ciencies Fotoniques, The Barcelona Institute of Technology, Mediterranean Technology Park, 08860 Castelldefels (Barcelona), Spain}
\author{Hugues de Riedmatten}
\affiliation{ICFO-Institut de Ciencies Fotoniques, The Barcelona Institute of Technology, Mediterranean Technology Park, 08860 Castelldefels (Barcelona), Spain}
\affiliation{ICREA-Instituci\'{o} Catalana de Recerca i Estudis Avan\c cats, 08015 Barcelona, Spain}

\date{\today}

\begin{abstract}
We report on the quantum storage of a heralded frequency-multiplexed single photon in an integrated laser-written rare-earth doped waveguide. The single photon contains 15 discrete frequency modes separated by 261 MHz and spaning across 4 GHz. It is obtained from a non-degenerate photon pair created via cavity-enhanced spontaneous down conversion, where the heralding photon is at telecom wavelength and the heralded photon is at 606 nm. The frequency-multimode photon is stored in a praseodymium-doped waveguide using the atomic frequency comb (AFC) scheme, by creating multiple combs within the inhomogeneous broadening of the crystal. Thanks to the intrinsic temporal multimodality of the AFC scheme, each spectral bin includes 9 temporal modes, such that the total number of stored modes is about 130. We demonstrate that the storage preserves the non-classical properties of the single photon, and its normalized frequency spectrum. 
\end{abstract}

\maketitle

Quantum memories for light are important devices in quantum information science \cite{Sangouard2011,Afzelius2015}. Multimode quantum memories would greatly help the scaling of quantum networks by decreasing the entanglement distribution time between remote quantum nodes \cite{Simon2007,Sangouard2011}. Current research focuses mostly on time multiplexing in rare-earth doped crystals \cite{Usmani2010,Clausen2011,Saglamyurek2011,Gundogan2015,Tiranov2016,Jobez2016,Seri2017,Kutluer2017,Laplane2017,Yang2018} and in spatial multiplexing in atomic gases \cite{Lan2009,Nicolas2014,Ding2015,Pu2017,Chrapkiewicz2017,Tian2017}.  Beyond these demonstrations, rare-earth doped crystals, thanks to their large inhomogeneous broadening, represent a unique quantum system which could also add another degree of freedom for multiplexing, i.e. the storage of multiple frequency modes \cite{Sinclair2014}. This unique ability could also enable the generation of  high-dimensional frequency entanglement between a photon and a matter system. Furthermore, it could also provide a quantum memory for frequency bin-encoded qubits, which are gaining interest both in quantum information and computation \cite{OlislagerCusseyNguyenEtAl2010, Lukens17,Riel2017,Kues2017, Lu2018}.  Here, we report on the first demonstration of quantum storage of a frequency-multiplexed single photon into a laser-written waveguide integrated in a \Prs crystal. The multimode capability of our memory is further increased thanks to the intrinsic temporal multimodality of the storage protocol used, i.e. the atomic frequency comb. 
The great advantage of using waveguides is that the power required to prepare the quantum memory is strongly reduced due to the increased light-matter interaction. This enables simultaneous preparation of several memories at different frequencies, with a moderate laser power.

Very few experiments have explored the storage of frequency-multiplexed photonic states. Qubits encoded with weak coherent states have been stored in up to 26 frequency modes in the excited state of a Tm-doped waveguide \cite{Sinclair2014}, and up to two modes in the spin state of a Pr-doped crystal \cite{Yang2018}. Parts of the spectrum of a broadband single photon were also stored in up to 6 frequency bins in an Er-doped optical fiber \cite{Saglamyurek2016}. 
In contrast, our source naturally generates photon pairs in discrete frequency bins \cite{Seri2018} that can all be stored in our crystal.  

In this paper, we demonstrate the storage of a frequency-multiplexed heralded single photon, consisting of about \numspmodes modes, in a Pr\3-doped waveguide \cite{Seri2018}. This leads to an increase of our count rates by 5.5 with respect to the single frequency mode storage that allows us to make a detailed analysis of the multiplexed biphoton state after the storage. We demonstrate the non classicality of the correlations of the multiplexed biphoton after a preprogrammed storage time of 3.5 $\mu s$. We also show that the normalized spectrum of our single photon is well preserved during the storage.

\bigskip

\begin{figure*}
	\centering{\includegraphics[width=2\columnwidth]{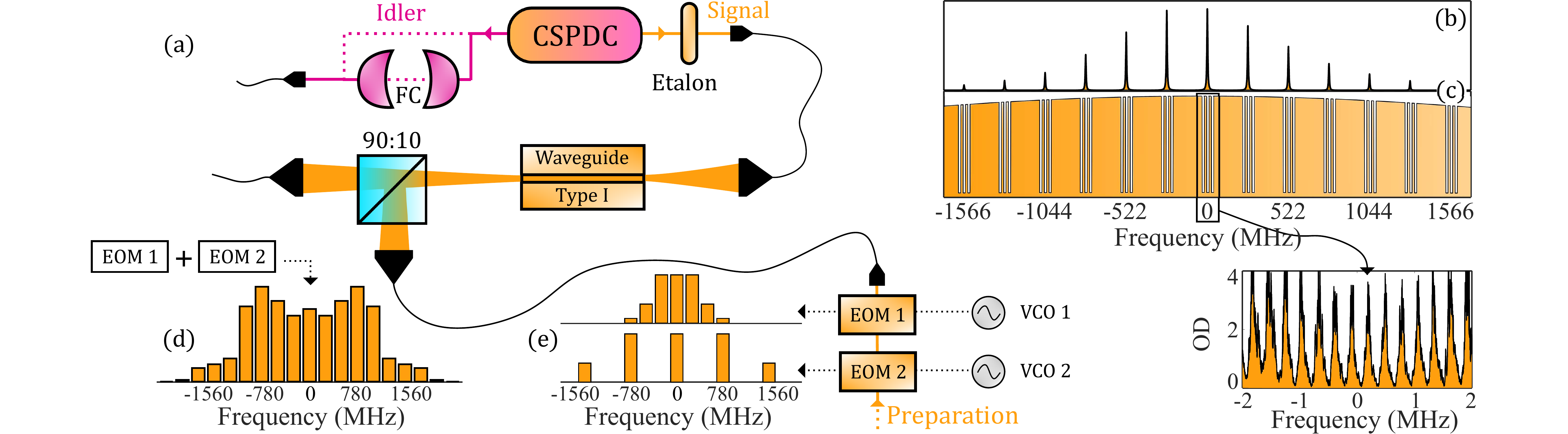}}
	\caption{(a) Experimental setup. From the photon-pair source (CSPDC) the idler photon (purple beam) is coupled to a fiber, either directly or after passing through a filter cavity (FC). The signal photon (orange beam) is sent to the memory crystal (waveguide type I). The preparation light (bottom) is modulated with two EOMs: EOM$_1$ (EOM$_2$) is driven with a voltage controlled oscillator VCO$_1$ @261.1 MHz (VCO$_2$ @3x261.1 MHz). A 90:10 (T:R) beam splitter couples $10\%$ of the preparation beam into the waveguide, counter-propagating with the signal photons. The modulation given by each EOM is in (e), the resulting one in (d). (b) Sketch of the generated bi-photon spectrum. (c) Sketch of the inhomogeneous broadening, tailored with many atomic frequency combs (AFCs). The inset is a trace of a measured AFC ($\tau=3.5\,\mu s$, measured efficiency $\eta_{AFC} = 8.5\%$) in optical depth (OD).} 
\label{fig1}
\end{figure*}

The setup, Fig. \ref{fig1}a, consists of two main parts: the photon-pair source and the memory. The first is based on cavity-enhanced spontaneous parametric down conversion (CSPDC) \cite{Rielander2014}. The idler photon, at 1436 nm (telecom E-band), heralds the presence of a signal photon resonant with an optical transition of Pr\3 at 606 nm. Around the SPDC source, a cavity resonant both with the idler and the signal frequencies redistributes the bi-photon spectrum along narrow frequency modes (Fig. \ref{fig1}b). These modes are separated by the cavity free spectral range (FSR = 261.1 MHz) and have a linewidth of 1.8 MHz. Because the signal and idler photons have different FSR, only a few modes (15 in our case) are generated \cite{Pomarico2012}. So far, only the central frequency mode (at frequency 0 in Fig. \ref{fig1}b) was heralded and stored \cite{Seri2017, Seri2018}. For most of the measurements reported here, the idler photon is directly sent to a single photon detector (SPD), resulting in a frequency-multimode heralding (MM\iin). Occasionally, a home-made Fabry-Perot filter cavity (FC in Fig. \ref{fig1}a) is used to provide a single frequency heralding (SM\iin).

\begin{figure*}
\centerline{\includegraphics[width=2\columnwidth]{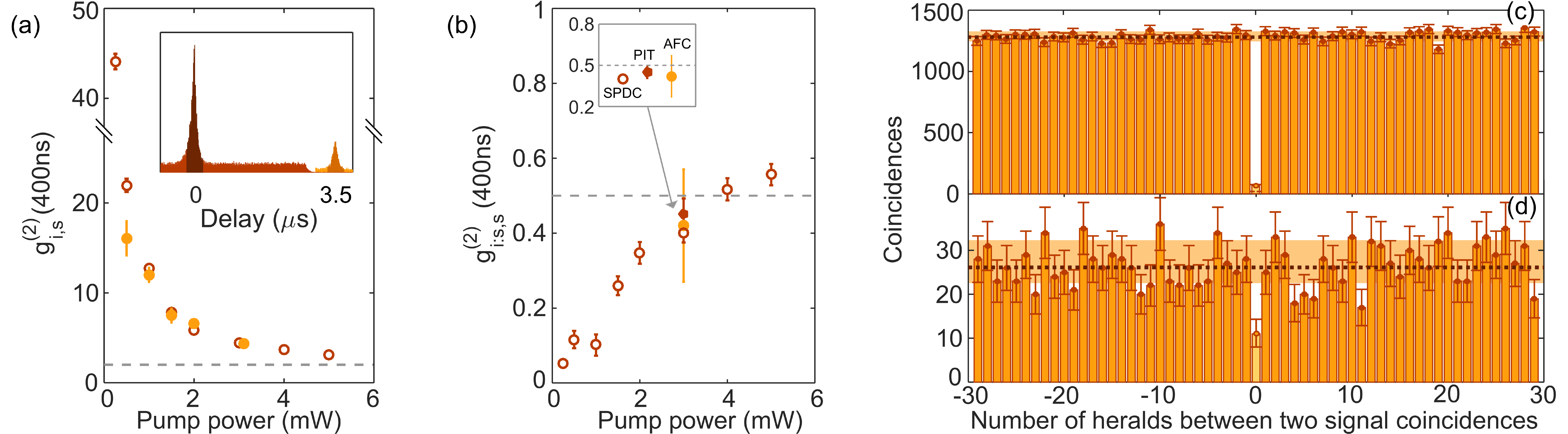}}
\caption{(a) Second order cross-correlation value $g^{(2)}_{s,i}(\Delta t)$ vs pump power (P) for the photon-pair source alone (empty brown circles) and after the storageof the signal (plain orange points). The inset shows a temporal coincidence histogram between signal and idler photons at P=3 mW (the input photon in brown and AFC echo in orange). (b) Conditional autocorrelation of the signal photon  $g^{(2)}_{i:s,s}(400ns)$ vs P for the source alone (empty brown circles). $g^{(2)}_{i:s,s}(400ns)$ for the signal photon transmitted through a spectral pit (plain brown dots) or after storage (plain orange points) are also shown for P=3 mW. (c) Autocorrelation histogram for the source alone for P=0.25 mW. (d) Autocorrelation histogram of the stored signal photons for P=3 mW (measurement time  $>44$ hours).}
\label{fig2}
\end{figure*}

The signal photon, filtered with an etalon, is coupled into a SM fiber and sent to the memory optical table. We store the signal photons in a 0.05\% doped \Prs crystal at cryogenic temperature ($<$3 K), where a laser-written type I waveguide was fabricated by femtosecond laser micromachining \cite{Seri2018}. If we want the photons to be transmitted through the waveguide, we produce a transparency window (spectral pit) around the signal frequency mode. The implemented storage protocol is the atomic frequency comb (AFC) technique \cite{Riedmatten2008a, Afzelius2009}: in this protocol the absorption line of the crystal is tailored into a periodical absorptive structure by optical pumping (inset of Fig.\ref{fig1}c). A photon absorbed by such structure is mapped into a collective superposition of atomic excitations. After a preprogrammed time, dependent on the comb periodicity, the atomic excitations rephase, releasing an echo of the incoming photon along the same direction. To store the whole spectrum of the frequency-multiplexed photons, we send the preparation beam to two cascaded electro-optical modulators (EOM$_{1,2}$) with resonance frequencies equal to 1$\times$FSR and 3$\times$FSR of the CSPDC source, respectively (see spectrum in Fig. \ref{fig1}d). With this method, using the same preparation time that we would need to prepare just one comb, we simultaneously tailor many AFCs (up to 15 demonstrated here, all for the same $\tau$) within the inhomogeneous absorption profile of Pr\3 (Fig. \ref{fig1}c, the inset shows a measured AFC trace). Hence, to perform multimode storage (MM\sn) we switch on both EOMs, while we switch them off to store just a single mode (SM\sn), i. e. to prepare the AFC only at the central frequency-mode. {We note that this method has a very favorable scalability, as the spectral modes stored scale exponentially with the number of EOMs used ($> 3^{\#\textrm{EOMs}}$).}

\bigskip

As a figure of merit for our photon pairs, we use the second-order normalized cross-correlation function between signal and idler fields, defined as $g^{(2)}_{s,i}(\Delta t) = p_{s,i}/(p_s p_i)$, where $p_{s,i}$ is the probability to detect a coincidence between the two and $p_s$ ($p_i$) the probability to detect independently a signal (idler) photon in a time window $\Delta t$. 
We measured \gcc vs pump power (P) for the heralded photons before and after the storage (brown empty and orange full dots in Fig. \ref{fig2}a, respectively). Here, both the idler and signal photons are multimode (MM\ii - MM\sn). 
An example of a trace from which we extract the \gcc is shown in the inset of Fig. \ref{fig2}a: the brown and orange traces represent coincidence histograms between idler and signal photons (P=3 mW) measured before the memory setup and after the storage, respectively. The time window considered, $\Delta t=400$ ns, is shown as a darker region in both the traces. {Thanks to the intrinsic temporal multimodality of the AFC scheme \cite{Afzelius2009}, we store $3.5\,\mu s/400\,ns\sim9$ distinguishable temporal modes}. As expected for a two-mode squeezed state \cite{Sekatski2012}, the \gcc increases while decreasing P (Fig. \ref{fig2}a). All the data points are above the classical limit (gray dotted line), defined by the Cauchy-Schwarz inequality as $\sqrt{g^{(2)}_{i,i}\cdot g^{(2)}_{s,s} }$ (for simplicity we overestimate a classical limit of 2). 
Except for the point at 0.5 mW, where the coincidence rate becomes comparable to the noise rate, the \gcc after the storage follows the same behaviour of the measurement with the source only. 

To fully characterize the non-classicality of the bi-photon, we measured the heralded autocorrelation of the signal photon \hac, namely the autocorrelation of the signal photon conditioned on an idler detection \cite{Fasel2004}. Here again, both the heralding and the signal photons are multimode (MM\ii - MM\sn). The signal photons are split by a fiber beam splitter and detected with two SPDs. A \hac histogram measured before the memory setup for P=0.25 mW is reported in Fig. \ref{fig2}c.  The \hac has been measured before the memory setup for different P, the lowest point (extracted from Fig. \ref{fig2}c) being $g^{(2)}_{i:s,s}$(400~ns)~$=0.052\pm 0.007$, deep in the single photon regime. To overcome the low statistics, we repeated the measurement of the \hac for  P=3 mW sending the signal photons to the memory setup, and preparing a multimode spectral pit or AFC. The comparison of the results for the same pump power is shown in the inset of Fig. \ref{fig2}b. 
All the measured data points of Fig. \ref{fig2}b are below the classical limit of 1. Moreover the \hac measured at 3~mW after the spectral pit ($g^{(2)}_{i:s,s}$(400~ns)~$=0.45\pm 0.04$) or after the storage in the AFC ($g^{(2)}_{i:s,s}$(400~ns)~$=0.42\pm 0.15$) are both compatible with the measurement performed before the memory setup ($g^{(2)}_{i:s,s}$(400~ns)~$=0.40\pm 0.03$). We thus conclude that the storage in the memory does not degrade the statistics of the single photons. 

\bigskip

\begin{figure}[htb]
\centering{\includegraphics[width=1\columnwidth]{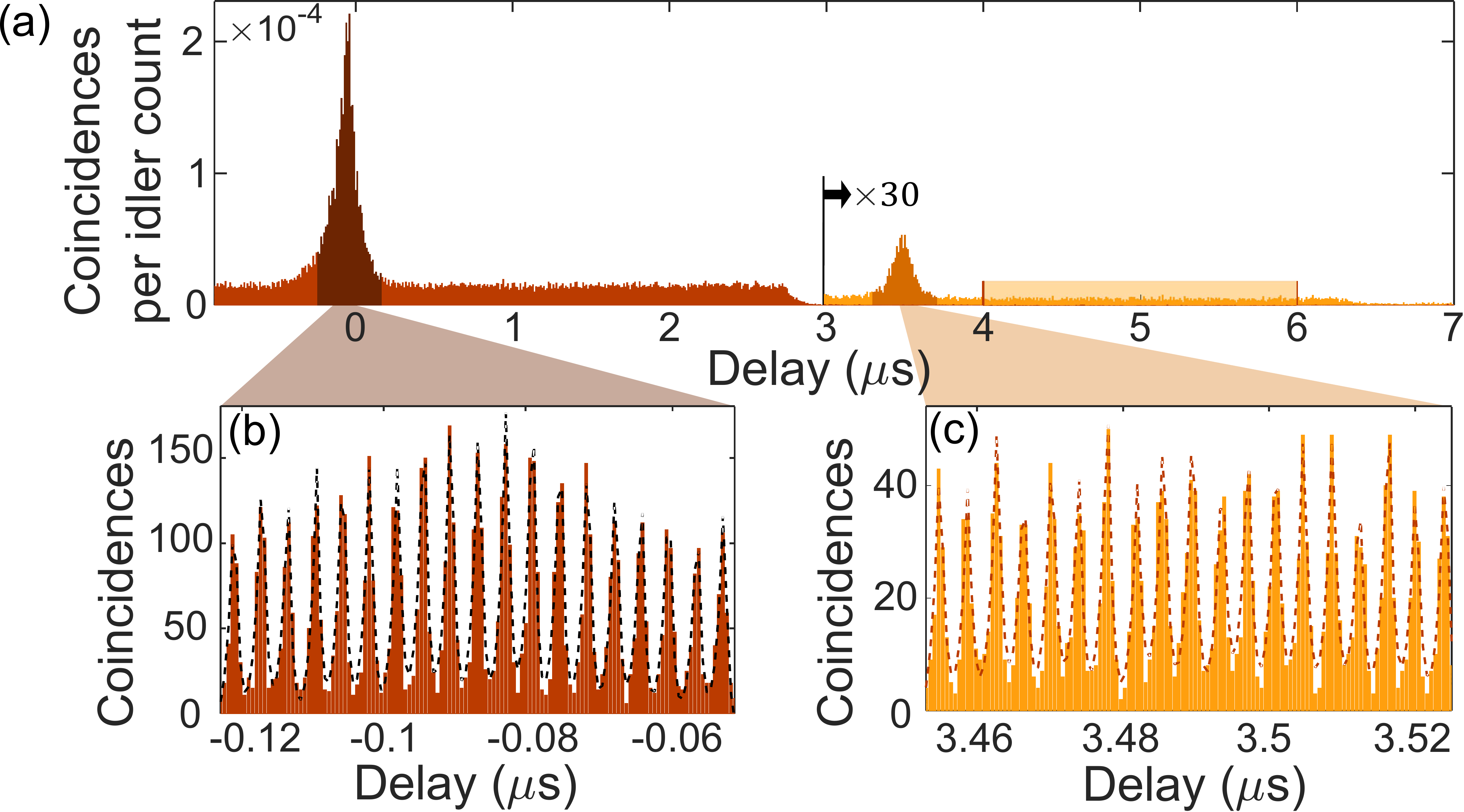}}
\caption{(a) Temporal coincidence histogram between idler and signal photons after the source (brown) or after storage (orange). The rectangle between 4 and 6 $\mu s$ is the stored noise, i.e. the accidental counts $p_s p_i$ in the \gcc of the AFC echo (the accidental counts for the input are measured before the brown peak). Panels (b) and (c) show a zoom in the peaks, where the beating between the spectral modes is visible. The width of the beating-peaks extracted from the fits is 910 ps, which agrees with the 970 ps expected (see Appendix \ref{Ajitter}).}
\label{fig3}
\end{figure}

We now study the spectrum of our signal photons, before and after the storage.
A signature of the presence of different frequency modes is the beating between them in the temporal \gcc function (Fig. \ref{fig3}a) \cite{Fekete2013}. The expected beating has a periodicity of $\sim3.8$ ns (1/FSR). We then recalculate the histograms with increased resolution (Fig.~\ref{fig3}b before the storage, Fig. \ref{fig3}c after). The clear beating in the AFC echo, with the inferred periodicity, is a strong signature that the storage protocol preserves the frequency multimodality of the photon. We expect the width of the oscillation peaks to decrease with the number of spectral modes interfering \cite{Fekete2013}, however, the time jitter of our detection system allows us only to infer a lower bound of 4 (see Appendix \ref{Ajitter}).

\bigskip

\begin{figure}[htb]
	\centering{\includegraphics[width=1\columnwidth]{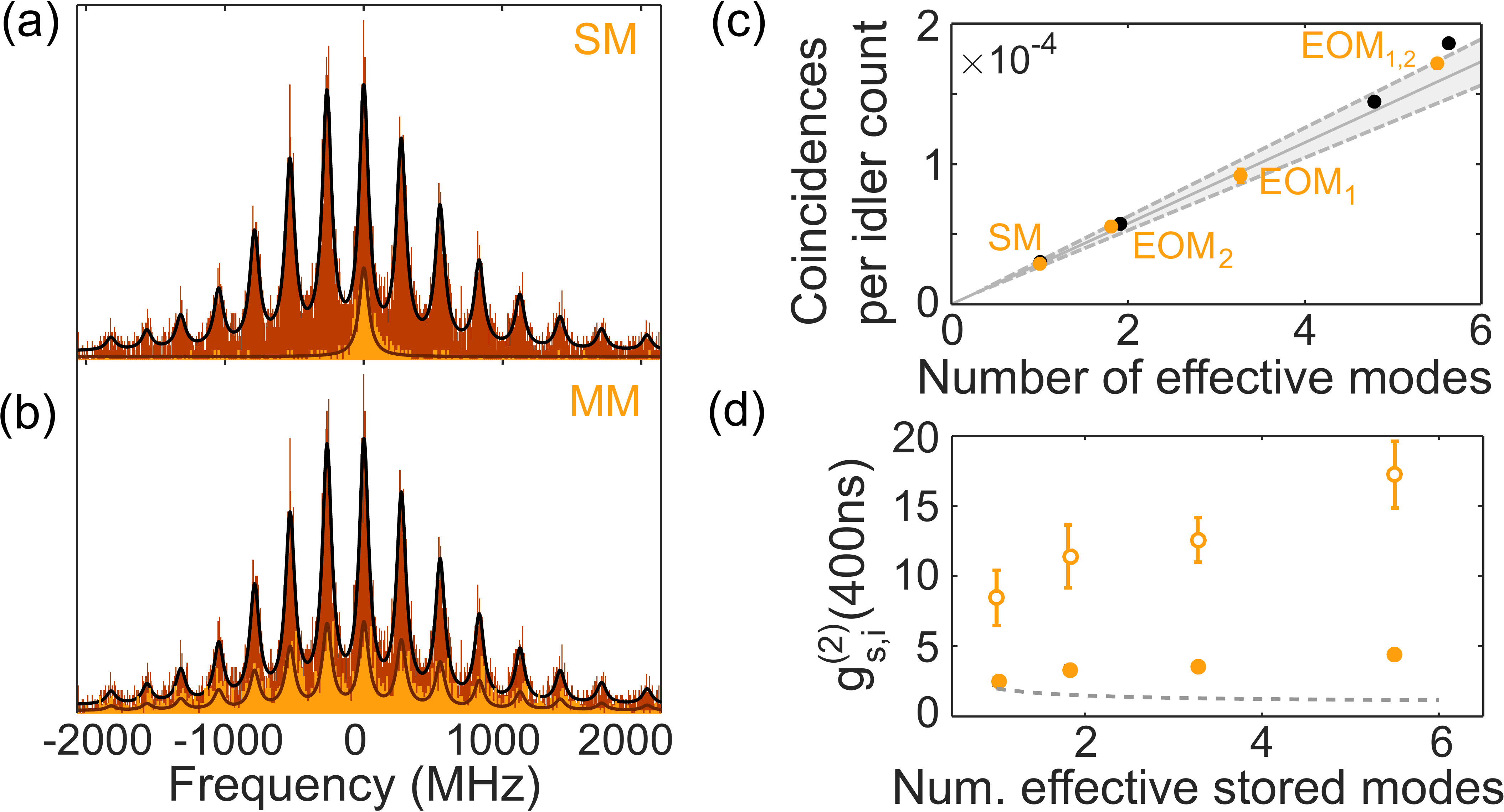}}
	\caption{(a)-(b) Heralded signal detections triggered to the scan of the FC in the case of SM and MM storage, respectively. The brown (orange) histogram is measured with the source alone (after the storage). The solid lines are fits to the brown histogram, renormalized to overlap with the orange trace. (c) Normalized coincidences between idler detections and signal photons, either after being transmitted through a spectral pit (black dots) or being stored (orange dots), versus number of effective modes N$_{eM}$. 
The gray line is fixed to pass through the origin and the first orange point (N$_{eM}=1$), the gray area being the error. The coincidence rate after the spectral pit is multiplied by the storage efficiency ($\eta_{AFC}=8.5\%$), to compare all the results with the same line. (d) \gcc measured versus number of effective stored modes. The orange full (empty) points are measured with P=3 mW (1 mW). The gray dotted  line is the theoretical classical limit, which scales with the number of modes N as 1+1/N (the real bound would be lower, due to noise and a finite detection window of 400 ns).}
	\label{fig4}
\end{figure}

To quantify the number of generated and stored frequency modes, we send the idler photons into the FC (Fig. \ref{fig1}a) that we scan about 4 GHz, covering the whole spectrum of the idler photons. We thus herald different signal frequency modes at different times. The coincidences detected before and after the storage are plotted in Figs. \ref{fig4}a and \ref{fig4}b  in brown and orange, respectively. In the SM\s case, we just detect coincidences when the FC is in resonance with the stored mode (Fig. \ref{fig4}a); if we switch on the two EOMs (MM\sn), we can store and retrieve the whole spectrum of the signal photon (Fig. \ref{fig4}b), i.e. about 15 spectral modes. As the spectrum of the preparation beam is almost flat for the nine central modes (Fig. \ref{fig1}d), the biggest part of the spectrum is stored with the same efficiency of the SM\s case. Despite the lower preparation power, also the other modes are stored with comparable efficiencies. We quantify the similarity of the normalized spectrum of the heralded photon before and after the storage from the measurements shown in Fig. \ref{fig4}b. We extract the count rate and the \gcc of each frequency mode and define an overlap function ranging from 0 (no overlap) to 1 (perfect overlap). We find an overlap between the count rates of $0.97\pm0.03$ and of $0.98\pm0.06$ between the \gcc values. More details on this analysis can be found in Appendix \ref{spectrumoverlap}.

\bigskip

As the spectrum of our bi-photon is not flat (Fig.~\ref{fig1}b), the count rate does not increase linearly with the number of stored modes. We define an \emph{effective mode} as a mode whose count rate is the same as the one of the central frequency mode. Hence, if we store just the central mode (SM\sn), we store 1 effective mode. On the other hand, the whole spectrum of the photon (15 spectral modes) will be equivalent to 5.6 effective modes. By using different EOM configurations, we can vary the number of effective modes stored (N$_{eM}$, see Appendix \ref{afc_pit_calib}). The light orange data points in Fig. \ref{fig4}c show the measured coincidence rate in the retrieved AFC echo versus N$_{eM}$. For this measurement the idler photon is not filtered (MM\iin). 
When increasing N$_{eM}$, the coincidence rate increases linearly, as expected. The same measurement is performed by preparing a multimode spectral pit (black dots in  Fig. \ref{fig4}c). For each EOM configuration considered in Fig. \ref{fig4}c, we also measure the cross-correlation of the AFC echo (full points of Fig. \ref{fig4}d). 
In this case, the \gcc increases with N$_{eM}$ because, even if both the coincidences and the stored noise increase linearly, there is a part of uncorrelated noise (due to dark counts or broadband noise), which remains constant (see Appendix \ref{scalingg2}). \\

The multimode storage technique demonstrated in this paper is directly suitable for storing the frequency bin entanglement naturally generated by our source \cite{Riel2017}. For that purpose it is desirable to reach higher values of \gcc. An easy way to increase the \gcc without destroying the entangled state is to reduce the pump power of the source (as shown in Fig. \ref{fig2}). We measure the \gcc versus N$_{eM}$, for P=1 mW (empty orange points in \ref{fig4}d). Despite the fact that the coincidence rate decreases together with P, the \gcc increases remarkably with respect to the previous measurement. The \gcc would increase further by filtering broadband noise before the idler detection with an etalon.

Another possible application of frequency multimode storage is to use each spectral channel as an independent quantum memory, leading to spectral multiplexing, e. g. for quantum repeaters \cite{Sinclair2014}. For this purpose we should distinguish the different heralding frequency modes, i.e. separate the modes of the idler photon \cite{Puigibert2017}. This would destroy the frequency bin entanglement, but the count rate and duty cycle of the experiment would increase, without decreasing the \gc. While it is challenging to  distinguish the different idler frequency modes with current technology due to their small separation,  we perform a series of measurements to validate our statement. We compare the \gcc measured with a single mode (SM$_i$) or multimode heralding photon (MM$_i$), storing either one mode (SM$_s$) or the whole spectrum (MM$_s$) of the signal photon (see table \ref{tabella}).

\begin{table}[H]
\centering
\begin{tabular}{|l l l l|}
\hline
SM$_s$ & MM$_i$ & $2.48\pm0.17$  &   \\ \hline
MM$_s$ & MM$_i$ & $4.35\pm0.22$ &   \\ \hline
MM$_s$ & SM$_i$ & $15.8\pm1.5$ &   \\ \hline
SM$_s$ & SM$_i$ & $72\pm12$ &   \\ 
\hline
\end{tabular}
\caption{Measured \gcc after storage in different configurations.}
\label{tabella}
\end{table} 
We note that, distinguishing the spectral modes of the heralding detections (SM$_i$), the \gcc of the retrieved multimode echo would increase. Moreover, if we could retrieve independently the different modes of the signal (fourth case), the \gcc for each spectral mode would be increased by a factor of 16, without decreasing the experiment count rate. 
We could also perform spin-wave storage \cite{Afzelius2010, Seri2017} addressing spin-states for each frequency mode individually,  e.g. using serrodyne frequency shifting of the control beams \cite{Johnson10}, having therefore a SM$_s$~-~SM$_i$ configuration.

\bigskip

We demonstrated quantum storage of a frequency multiplexed single photon, counting 15 spectral modes over 4 GHz, in an integrated rare-earth-doped laser-written waveguide.  Our work opens prospects for the realization of frequency multiplexed quantum repeaters, and for the demonstration of high dimensional frequency entanglement between light and matter. Together with the 9 temporal modes stored as an intrinsic property of the AFC protocol, we demonstrate the storage of more than 130 individual modes. Our results show that integrated waveguides in rare-earth doped crystals can be used as versatile light-matter interaction platforms with both time and frequency multiplexing capabilities. Moreover, the unique 3-dimensional fabrication capability of  laser-written waveguides \cite{Crespi2016} also holds promises for implementing large memory arrays in one crystal and allows fabrication of optical angular momentum compatible waveguides \cite{Chen2018}. The ability to combine several multiplexing capabilities in one system would open the door to the realization of massively multiplexed quantum memories.

\bigskip

\textbf{Acknowledgments.} We acknowledge financial support by the European Research Council (ERC) via the Advanced grant CAPABLE (742745), by the European Union via the Quantum Flagship project QIA (820445), by the Spanish Ministry of Economy and Competitiveness (MINECO) and Fondo Europeo de Desarrollo Regional (FEDER) (FIS2015-69535-R), by MINECO Severo Ochoa through grant SEV-2015-0522, by Fundaci\'o Cellex, and by CERCA Programme/Generalitat de Catalunya.

\section*{APPENDIX}
\label{app}

\subsection{Frequency multiplexed photon pair source}
To produce narrow single photons we embed a type-I periodically-poled lithium niobate (PPLN) nonlinear crystal inside an optical cavity (CSPDC). The effect of a cavity around the crystal is that only photons compatible with the resonator spectrum (linewidth 1.8 MHz) will be generated. In our case, we pump the CSPDC source with a cw laser at 426 nm producing a pair of photons at 606 nm (signal) and 1436 nm (idler) via spontaneous parametric down-conversion (SPDC). 

In order to enhance the generation of both photons, we need to lock our resonator at both wavelengths. Our lock method follows the Pound$-$Drever$-$Hall technique. To lock the cavity at the signal frequency, 606 nm, we send a reference laser at the same wavelength inside the cavity and we use the reflected light to generate an error signal. This is used as a feedback to a piezoelectric element installed in one of the mirrors of the cavity. We also use the light generated by difference frequency generation (DFG) inside the cavity to lock the cavity at the idler frequency. From this light we generate another error signal, used to know when the cavity is in resonance with a 1436 nm wavelength. We feed this error signal back to the current of the diode of the pump laser (tuning its frequency).

\begin{figure}[hbtp]
\centering
\includegraphics[width=.87\columnwidth]{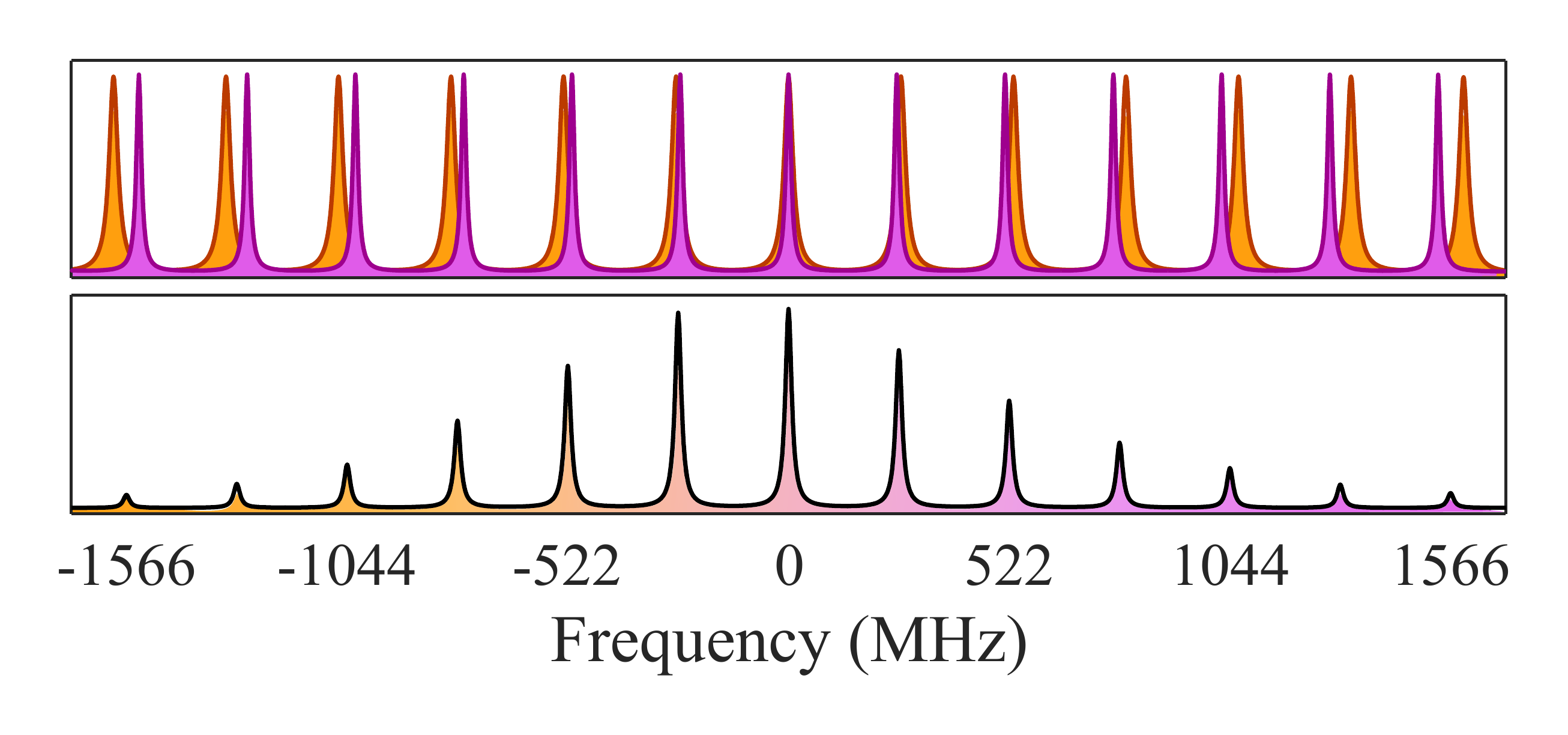}
\caption{Top: sketch of the spectral modes structure of the 606 nm photons (orange) and 1436 nm photons (purple) that have slightly different FSR inside our cavity. Bottom: sketch of the photon pairs generated inside the cavity, given by the overlap between signal and idler spectra. The linewidth of the individual modes was increased on purpose for the sake of clarity.
\label{cluster}}
\end{figure}

Having this double resonance with non degenerate photons has a fundamental implication for this work. The refractive index of the nonlinear crystal is slightly different for 606 nm and for 1436 nm. This implies that the free spectral range (FSR) of the two different wavelengths will be also slightly different (as sketched in Fig. \ref{cluster}). Due to the double lock of the cavity, only the pairs where both photons are resonant are extracted. This creates a clustering effect \cite{Pomarico2012}. The size of the cluster depends on the relation between the linewidth of the photons and the FSR. In our case the biphoton linewidth is measured to be 1.8 MHz and the FSR 261.1 MHz.

Measurements performed also in this work with a spectrum analyser (FC in Fig. 1 of the main text, linewidth 80 MHz, FSR 17 GHz) revealed that our source generates up to 15 spectral modes within the central cluster. 

For our measurements we switch off the pump laser of the CSPDC source triggered on the detections of idler photons to reduce broadband uncorrelated noise while the AFC echo is expected. Before sending the signal photon to the memory table, we filter it from broadband noise with an etalon (Fig. 1(a) of the main paper, linewidth 4.25 GHz, FSR 100 GHz)

\subsection{AFC preparation}
\label{afc_pit_calib}

Due to the vibrations given by the cryostat, the light is coupled into the waveguide for about half of the cryostat cycle. 

Differently from our previous realizations, the preparation beam for the AFC is counter-propagating with respect to the single photons. It is coupled into the waveguide after being reflected by a 90:10 (T:R) BS (see experimental set up in Fig. 1 of the main text). In this way the two paths are independent, so that we can couple into the waveguide the preparation light during one phase of the cryostat cycle and the single photons during a different phase of the cycle. By doing this, we use the whole cryostat cycle and, as a direct consequence, we gain time for both measuring single photons and for the comb preparation.

\subsubsection{AFC efficiency and spectral pit transmission for different spectral modes}

\begin{figure}[hbtp]
\centering
\includegraphics[width=1\columnwidth]{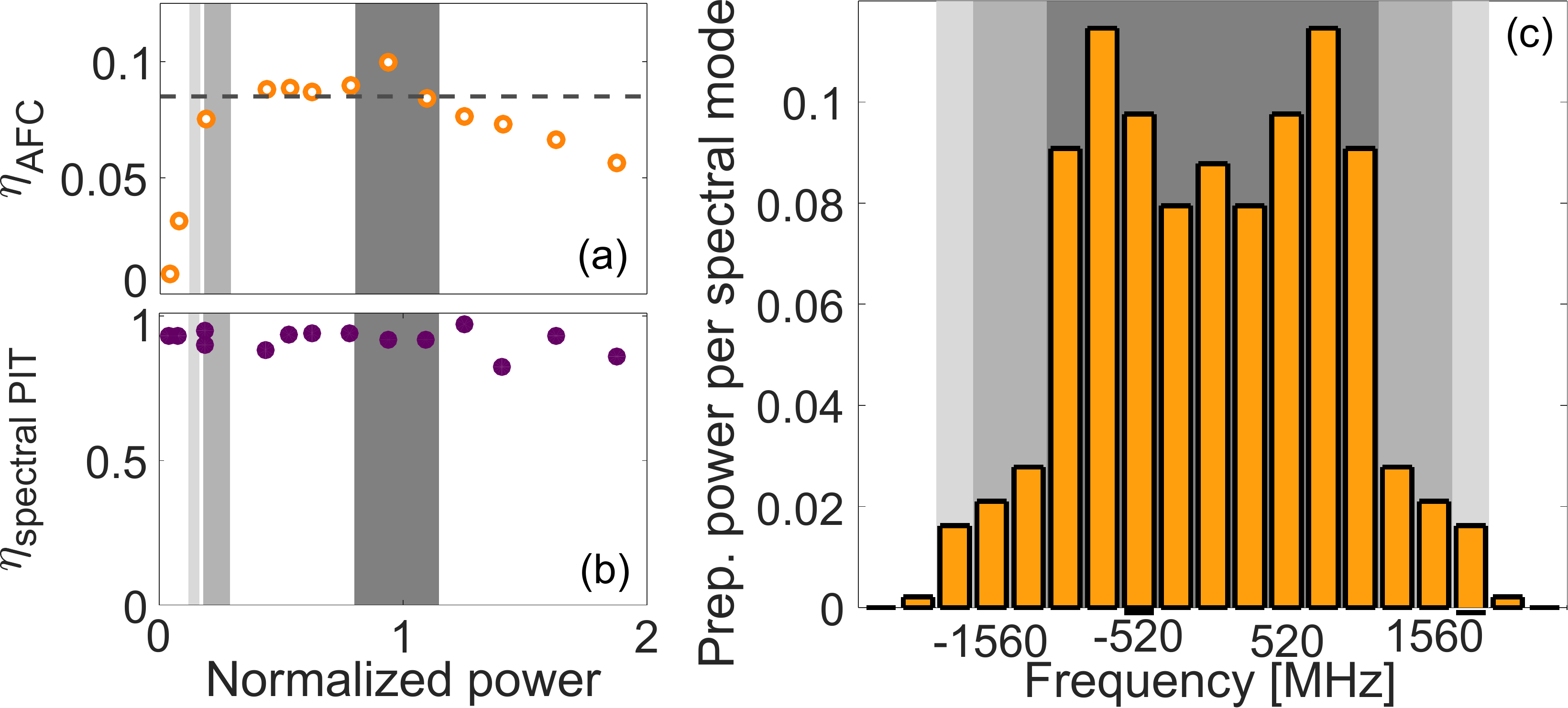}
\caption{Calibration measurements to estimate the expected efficiency of the pit and the AFC for the different spectral modes. (a) Dependence of the AFC efficiency on the power that we use to prepare it. (b) Dependence of the pit efficiency on the power that we use to prepare it. For (a) and (b) in the x axis a value of 1 corresponds to the optimum value of power to send to the waveguide. (c) Amplitude (normalized to the input power) of the spectral modes of the preparation light after the two EOMs. Gray rectangles are used to guide the reader to the expected efficiency for each mode.\label{fig:afc_pit_calib}}
\end{figure}

As we do not access independently the different spectral modes of the signal photons, the storage efficiency and the transmission through the spectral pit are not measured directly for each of them. Our calibration will in fact involve only the central mode: we prepare the AFC and spectral pit, in a single mode configuration (EOM$_{1,2}$ off), for different preparation powers. The results are shown in Figs. \ref{fig:afc_pit_calib}(a) and \ref{fig:afc_pit_calib}(b), respectively. The power that we want to have, for each of the 9 central spectral modes of the preparation, is 160 $\mu$W before the waveguide (this is what we call 1 in the x-axis of the figures). For this power, in a single mode configuration, we measured an efficiency $\eta_{AFC}=8.5\%$, reported as dotted line in Fig. \ref{fig:afc_pit_calib}(a).  

When we switch on the EOMs, the preparation power is shared between many spectral modes. The power in each mode, for the decided configuration of our experiment, is described in figure \ref{fig:afc_pit_calib}(c). From this calibration, we can infer the expected storage efficiency and transmission through the spectral pit for the central 9 modes (dark gray ragion in Fig. \ref{fig:afc_pit_calib}), the modes $\pm 5$ and $\pm 6$ (gray region) and the modes $\pm 7$ (light gray region). We note that the efficiency of the AFC seems to get visibly lower only for the modes $\pm 7$ (Fig.  \ref{fig:afc_pit_calib}(a)), while the spectral pit does not worsen while decreasing the power (in the range of powers analyzed).

An equivalent calibration was performed for the cases of only EOM$_1$ on or EOM$_2$ on: in the single EOM case, the central 5 modes are stored with almost flat efficiency; the modes $\pm 3$, where the preparation power is about $10\%$ of the central ones, are also stored, but with a much lower expected efficiency ($\sim 3.5\%$).

\subsubsection{Number of Effective Modes, N$_{eM}$}

As the spectrum of our bi-photon is not flat, the count rate does not increase linearly with the number of stored modes. Hence, for the analysis of Fig. 4c and d of the main text, we define a quantity that we call \emph{effective mode}, which is a mode whose count rate is the same as the one of the central frequency mode. If we store just the central mode, we store 1 effective mode. On the other hand, the whole spectrum of the photon (15 spectral modes) is equivalent to 5.6 effective modes.

By using different EOM configurations, we can vary the number of effective modes stored, N$_{eM}$. This quantity is calculated starting from the spectrum of the photon and of the preparation beam in all the EOMs configurations. Knowing these quantities, we estimate the storage efficiency or transmission through the spectral pit for each mode, using the method described in the previous subsection. For example, in Fig. 4(c) of the main text, the points, from left to right, are measured with both EOMs off (N$_{eM} = 1$), EOM$_2$ on (N$_{eM} = 1.8$, as we store mainly the modes with low count rate), EOM$_1$ on (N$_{eM} = 3.3$, bigger than EOM$_2$ as we store the central part of the spectrum) and both EOMs on (N$_{eM} = 5.5$, different than the case of the only source because the modes $\pm7$ are stored with a slightly lower efficiency), respectively.

\subsection{Estimation of the number of frequency modes from the beating in the g$^{(2)}_{s,i}$}
\label{Ajitter}
\begin{figure}[hbtp]
\centering
\includegraphics[width=.65\columnwidth]{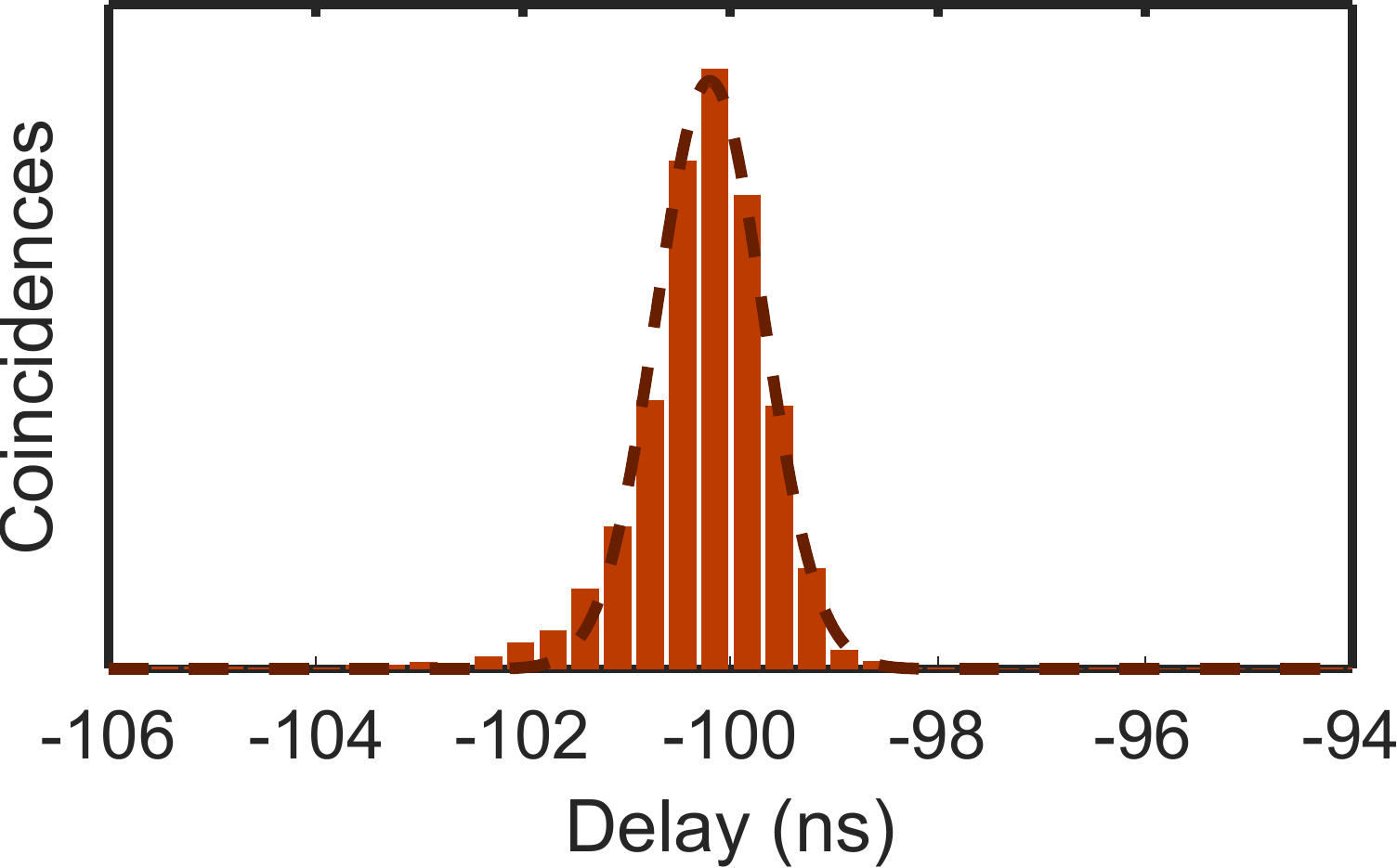}
\caption{Measured temporal jitter of our detection system. \label{jitter}}
\end{figure}

As explained in \cite{Fekete2013}, the width of the peaks in the beating measurement gives information about the number of modes that are coherently interfering. However, in our case, the time resolution required to perform this measurement is comparable to the temporal jitter of our detection system.

To measure the time resolution of our detectors, we measure photon pairs coming directly from a SPDC process in a similar non-linear crystal. Their bandwidth corresponds to hundreds of GHz. Therefore, the time width of the measured correlation peak will correspond to the convolution of all the experimental jitters that are affecting our measurements. This measurement (brown histogram) together with its fit to a Gaussian function (black dotted line) is shown in Fig. \ref{jitter}. From the fit we extract a jitter of 730 ps.

\begin{figure}[hbtp]
\centering
\includegraphics[width=.73\columnwidth]{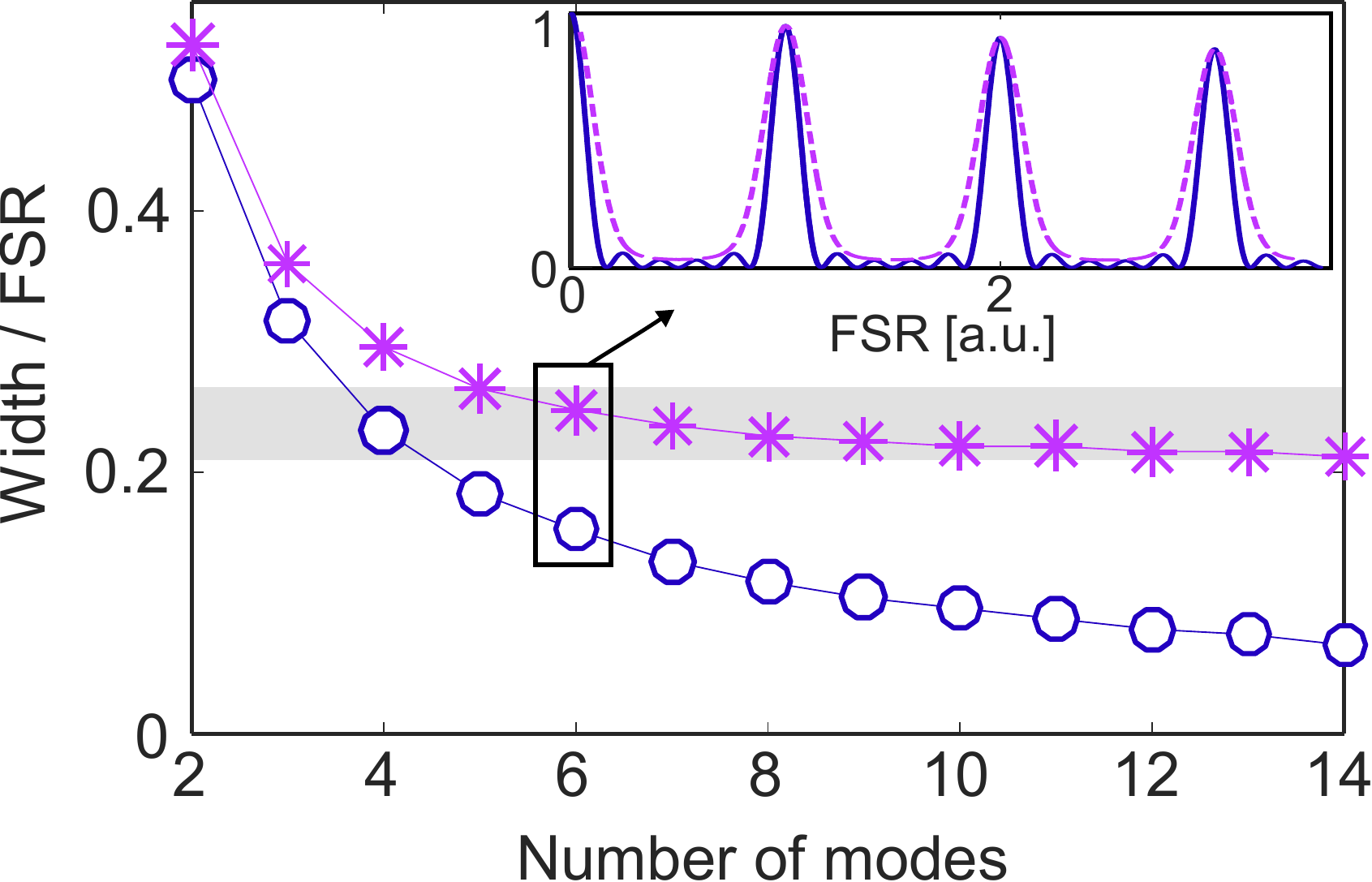}
\caption{In blue (empty circles), the theoretical prediction of the scaling of the ratio between the width of our beating and the FSR of our source versus the number of spectral modes interfering. In purple (stars), the expected behavior given our experimental jitter. The shaded area corresponds to the points compatible with the error of the fit done in Fig. 3b of the main text. The inset compares the simulated beatings for 6 spectral modes with (dotted line) and without jitter (solid line).}\label{JuliaConv}
\end{figure}

From the model explained in the Suppl. Material of \cite{Fekete2013}, reported as blue empty circles in figure \ref{JuliaConv} (solid line in the inset), we expect a width of 600 ps for a beating of 6 spectral modes (close to our effective stored modes). If we convolute the theoretical beating (blue solid line in the inset) with the measured jitter (Fig. \ref{jitter}), the expected width of the beating peaks increases, as can be seen in purple in Fig. \ref{JuliaConv} (stars and dotted line in the inset). The shaded area in Fig. \ref{JuliaConv} is the width of the beating peaks, $910 \pm110$ ps, extracted from the fit of Fig. 3b of the main text. From this measurement it is difficult to extract the number of spectral modes involved, but we can give a lower bound of 5 modes. Moreover, without making any consideration on the temporal jitter of our detection system, we can infer that the interfering modes are at least 4 (being our measurement compatible with the theoretical expectation for 4 modes, blue line of Fig. \ref{JuliaConv}). 

\begin{figure}[hbtp]
\centering
\includegraphics[width=1\columnwidth]{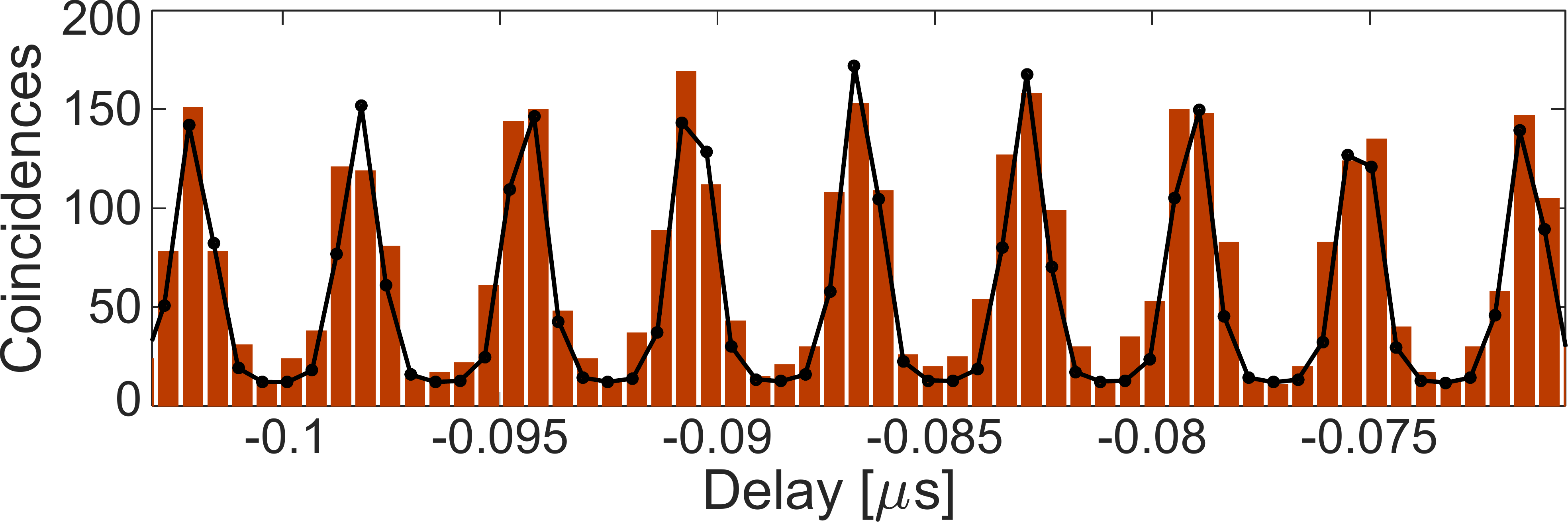}
\caption{Same histogram as the one of Fig. 3b of the main text. but the line corresponds to the simulated behavior expected for 6 spectral modes.}\label{beating_sim}
\end{figure}

Figure \ref{beating_sim} reports the same beating histogram shown in Fig. 3b in the main text. The black line is the expected beating for 6 modes, plotted with the same bin-size of the measured histogram (567 ps).

\subsection{Single mode and multimode comparison}

We show, in the main text, that the beating in the time resolved histogram is present both in the generated photon and in the AFC echo (Fig. 3 of the main text). This is a sign that after the storage of many frequency modes the coherent interference between them is maintained. In this section we check that the beating vanishes after the storage of only 1 frequency mode.

\begin{figure}[hbtp]
\centering
\includegraphics[width=.5\columnwidth]{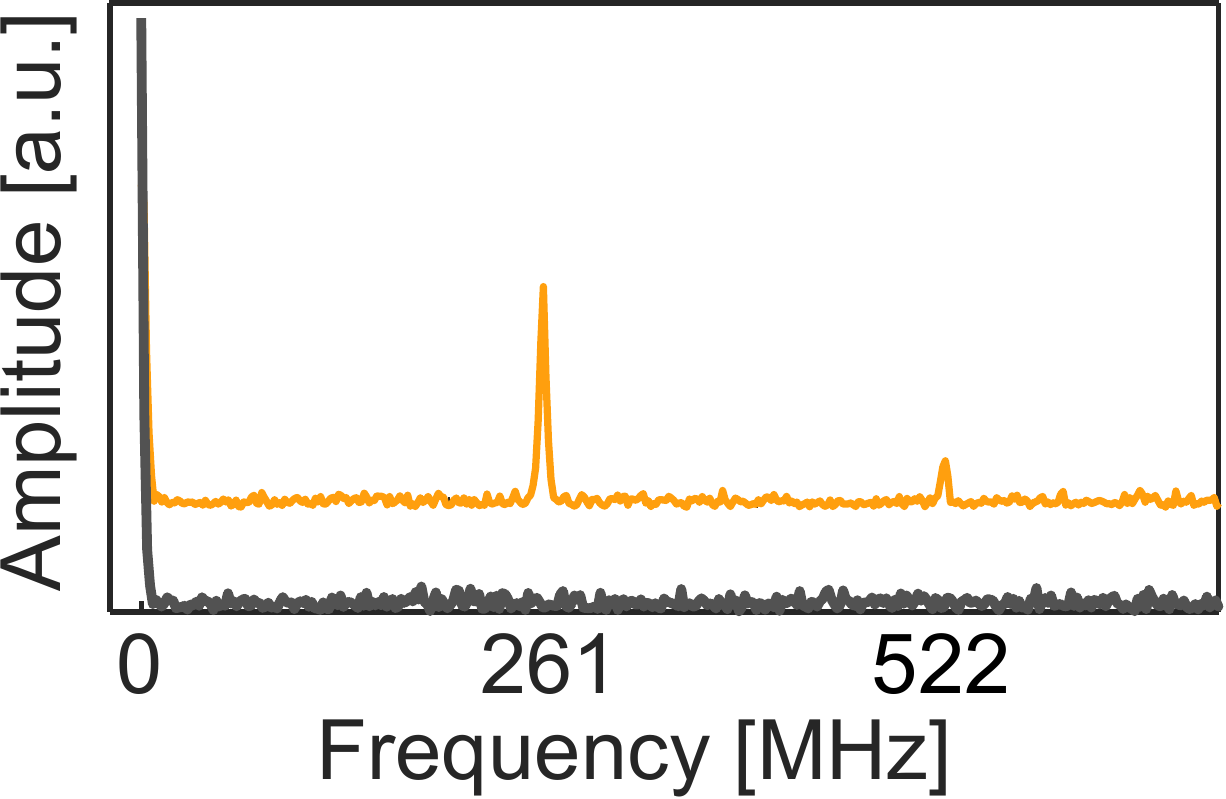}
\caption{Fourier transform of the correlations histogram around the AFC echo. The light orange line (shifted vertically for semplicity) corresponds to the case where we store all the spectral modes and the dark gray line to the case where we only store 1 spectral mode. \label{nobeating}}
\end{figure}

In Fig. \ref{nobeating} we report the Fourier transform of the coincidence histogram between the idler photon and the stored signal, around the correlations peak at 3.5 $\mu s$: in light orange, with an arbitrary vertical offset, for the case of multimode AFC preparation (MM$_s$), in dark gray for the storage of only 1 frequency mode (SM$_s$).

\subsection{Spectrum analysis of the heralded photons \label{spectrum}}

In this section we will explain how we build the histogram of Fig. 4(a) and 4(b) in the main text (and Fig. \ref{EOMcomb} of Appendix \ref{Aeomspectrumdiffcomb}).

To analyse the spectrum of our photon pair we use a Fabry-Perot (FP) cavity with 80 MHz of linewidth and a FSR of 17 GHz, installed in the idler optical path as can be seen in the Fig. 1 of the main text.

By scanning the FP cavity we are able to sweep in time the spectrum of the idler photons. In fact, at a specific time, only one specific frequency will be resonant with the cavity and thus detected. We post$-$select only the signal photons that are correlated with the detected idler photons. Consequently we are able to build a histogram by correlating the trigger of the scan with the heralded signal list. This histogram will correspond to the convolution between the spectrum of the photons and the spectrum of the FP cavity (a Lorentzian of 80 MHz of linewidth).

\begin{figure}[hbtp]
\centering
\includegraphics[scale=0.32]{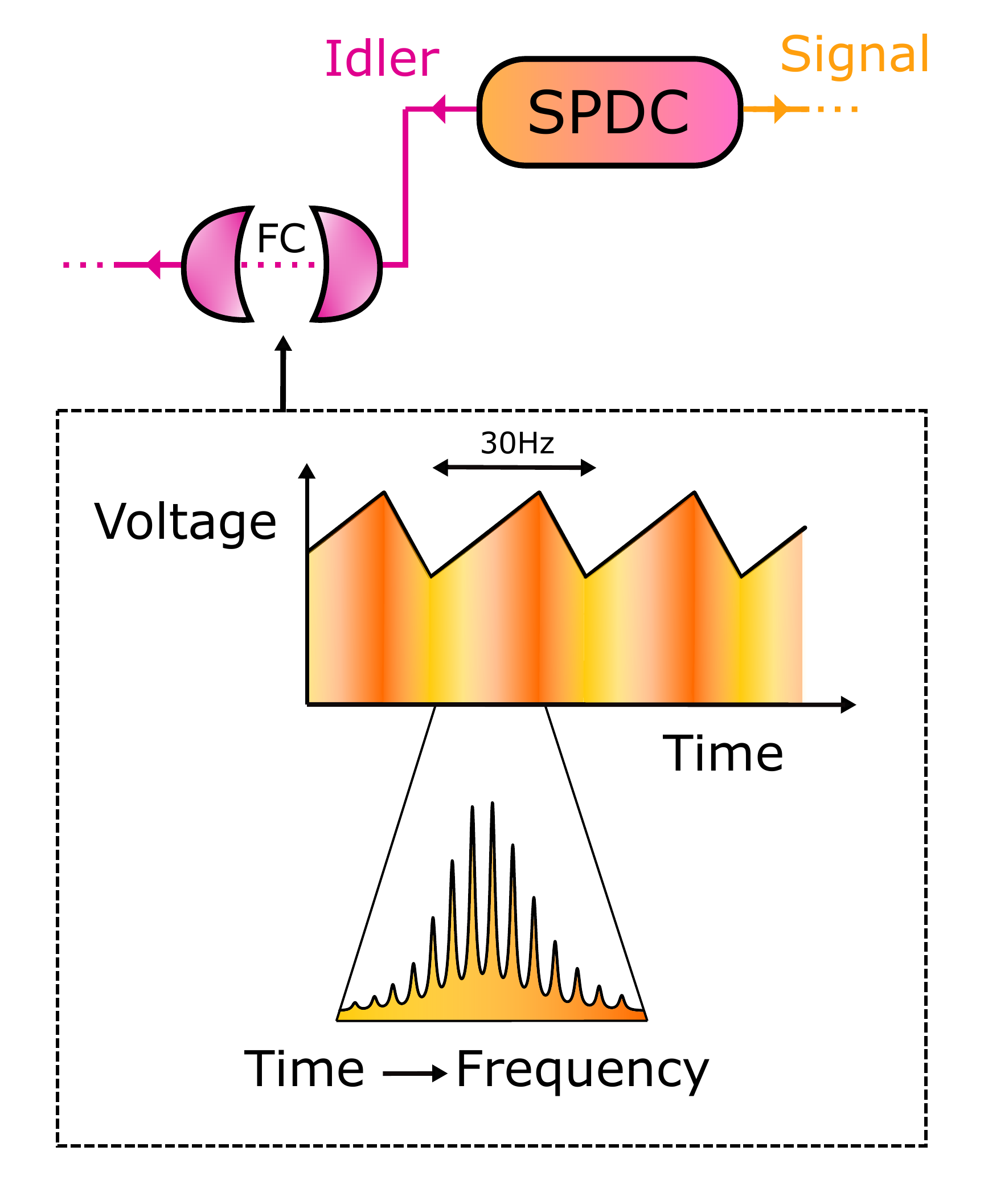}
\caption{Schematic of our spectrum analyser. \label{scan} }
\end{figure}

We scan a piezoelectric element installed in one of the mirrors of the FP cavity. We send an asymmetric ramp signal generated in a function generator that we map into a voltage ranging from 0 V to 150 V. The frequency of the scan is 30 Hz. We tune the offset of the signal in order to center the spectrum of our photons to the center of the slow part of the sweep (see Fig. \ref{scan}).

For all the measurements performed in this work, we scanned the FP cavity over $\sim4$ GHz, that should be enough to observe all the spectral modes of our photons.

\subsection{Fit to the spectrum of the CSPDC \label{fits}}
This section explains the fits that we used for Fig. 4a and 4b of the main text and for fig \ref{EOMcomb} of the SM.

We compare the spectrum of the stored frequency multiplexed heralded single photons with the photons right after the CSPDC source. For this propose, we fit the spectrum of the photons after the CSPDC source to a train of Lorentzian functions:
\begin{equation}
S = \sum_{i = 1}^{14} a_i \frac{\sigma/2}{(x-b_i)^2+(\sigma/2)^2} + d.
\end{equation}

Note that, since we use a Fabry-Perot cavity that is much broader than the photons to perform this measurement, it is fair the fit to Lorentzian functions (given that the spectral shape of a FP cavity follows a Lorentzian function).

For comparison with the spectrum after the storage, instead of doing a new fit we just re-scaled the original fit while preserving the same relative amplitudes.

\subsection{Post-processed analysis of the spectrum before and after the storage}
\label{spectrumoverlap}
\begin{figure}[hbtp]
\centering
\includegraphics[width=1\columnwidth]{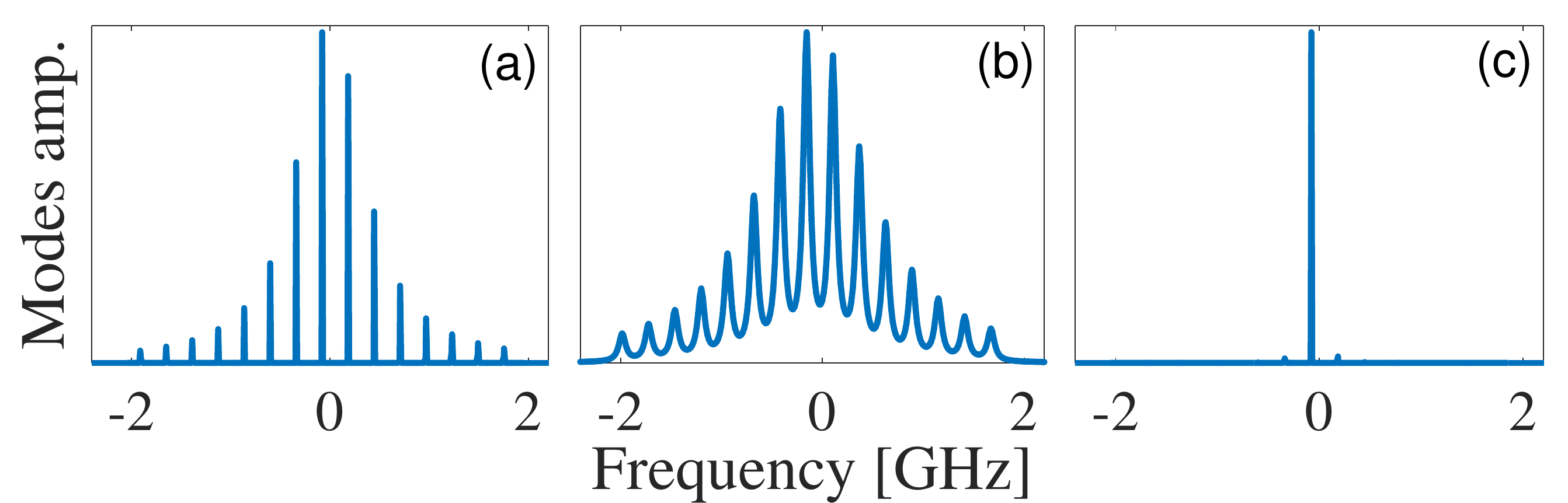}
\caption{(a) Frequency spectrum of the generated photons. (b)-(c) 
Simulated spectrum of the photons after a Fabry-Perot filter cavity (FC) with 80 MHz of linewidth, scanned along the spectrum or locked to the central mode, respectively.}
\label{spectrum_convol_lock}
\end{figure}

By scanning the idler spectrum with a filter cavity, we are in resonance at different times with different spectral modes. This means that at different times our idler detections herald different modes in the signal path. We plot a coincidence histogram between the tiggers and the idler detections. We then cut the idler list around each spectral mode and create one idler list for each spectral mode. In this way we want to post process a single mode idler case. 
Just for comparison we report in figure \ref{spectrum_convol_lock} the spectrum of the photon (a), the simulated spectrum scanning the cavity (b), and the spectrum that we would have by locking the cavity to the central mode (c). In figure \ref{spectrum_convol_lock}(b), the offset in the central mode is given only by the sum of the contribution of the other modes.

\begin{figure}[hbtp]
\centering
\includegraphics[width=1\columnwidth]{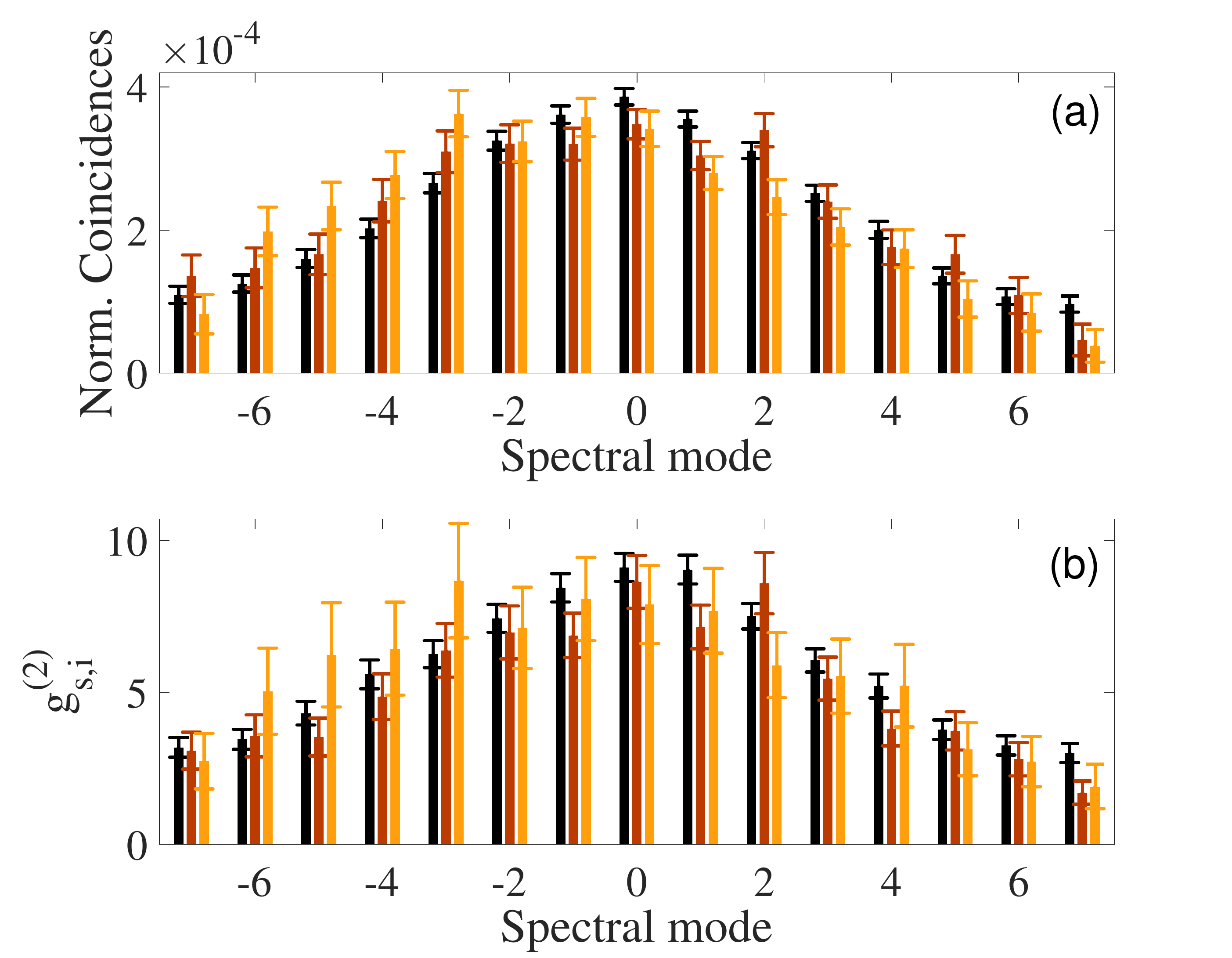}
\caption{(a) Relative amplitude of the count rates if we isolate the contribution of each spectral mode in post processing. The black and brown bars are renormalized to the average of the orange ones in order to better compare the spectrum. (b) Cross correlation of each spectral mode after post processing individually each of them. In both (a) and (b) the signal photons are measured right after the source (plotted in black), after passing through a spectral pit (brown bars) and after the AFC (orange bars).}
\label{fig:SMi_from_spectrum}
\end{figure}

Even if the analysis is not equivalent to having a single-mode configuration in the idler, due to the cross-talk between the modes, we can compare the coincidence rates (Fig. \ref{fig:SMi_from_spectrum}(a)) and \gcc (Fig. \ref{fig:SMi_from_spectrum}(b)), between the post-processed idler lists and the signals, in three different cases: just the source, black bars in the plots; sending the signals through spectral pits, brown bars; storing the signals with AFC, light orange bars. From both plots, it is possible to see that the shape of the spectrum is maintained after the storage.

We call CR$_{\textrm{CSPDC}}$ the coincidence rate between the signals and idlers with just the source (black bars in Fig. \ref{fig:SMi_from_spectrum}(a)) and CR$_{\textrm{AFC}}$ the coincidence rates measured after the storage (light orange bars). CR$_{\textrm{CSPDC}}$, in the figure, is normalized to the average CR$_{\textrm{AFC}}$ in order to compare the two spectra. Each of the two vectors is then normalized as: CR$_x$ = CR$_x$/$\sqrt{(\textrm{CR}_x\cdot \textrm{CR}_x)}$. We quantify the overlap between the spectrum before and after the storage by measuring the scalar product between the vectors CR$_{\textrm{CSPDC}}$ and CR$_{\textrm{AFC}}$. We find CR$_{\textrm{CSPDC}}\cdot$CR$_{\textrm{AFC}}=0.97\pm0.03$.

The overlap have been measured in the same way for the \gcc measurement, again with the \gcc between signal and idler measured with just the source (\gcc$_\textrm{CSPDC}$, black bars in Fig. \ref{fig:SMi_from_spectrum}(b)) and after the storage (\gcc$_\textrm{AFC}$, light orange bars in Fig. \ref{fig:SMi_from_spectrum}(b)). We find \gcc$_\textrm{CSPDC}\cdot$\gcc$_\textrm{AFC}=0.98\pm0.06$. As in the previous case, each vector is renormalized to the square root of its own scalar product.

In the paper we state that the \gcc expected in the case of a SM$_i$ for the central mode is $15.8\pm1.5$ (see table I of the main paper). This is actually the measured value with the idler passing through the FC locked around the central mode. As we discussed at the beginning of this section, when we scan the filter cavity, an important contribution around the frequency of the central mode is given by the other spectral modes. Taking those contributions into consideration, we expect a \gcc$= 9.8\pm1.1$ for the central mode in the case of just the source. This value matches with the measured value of $9.1\pm0.5$.


\subsection{Spectrum after storage with different EOM combinations}\label{Aeomspectrumdiffcomb}
\begin{figure}[hbtp]
\centering
\includegraphics[width=1\columnwidth]{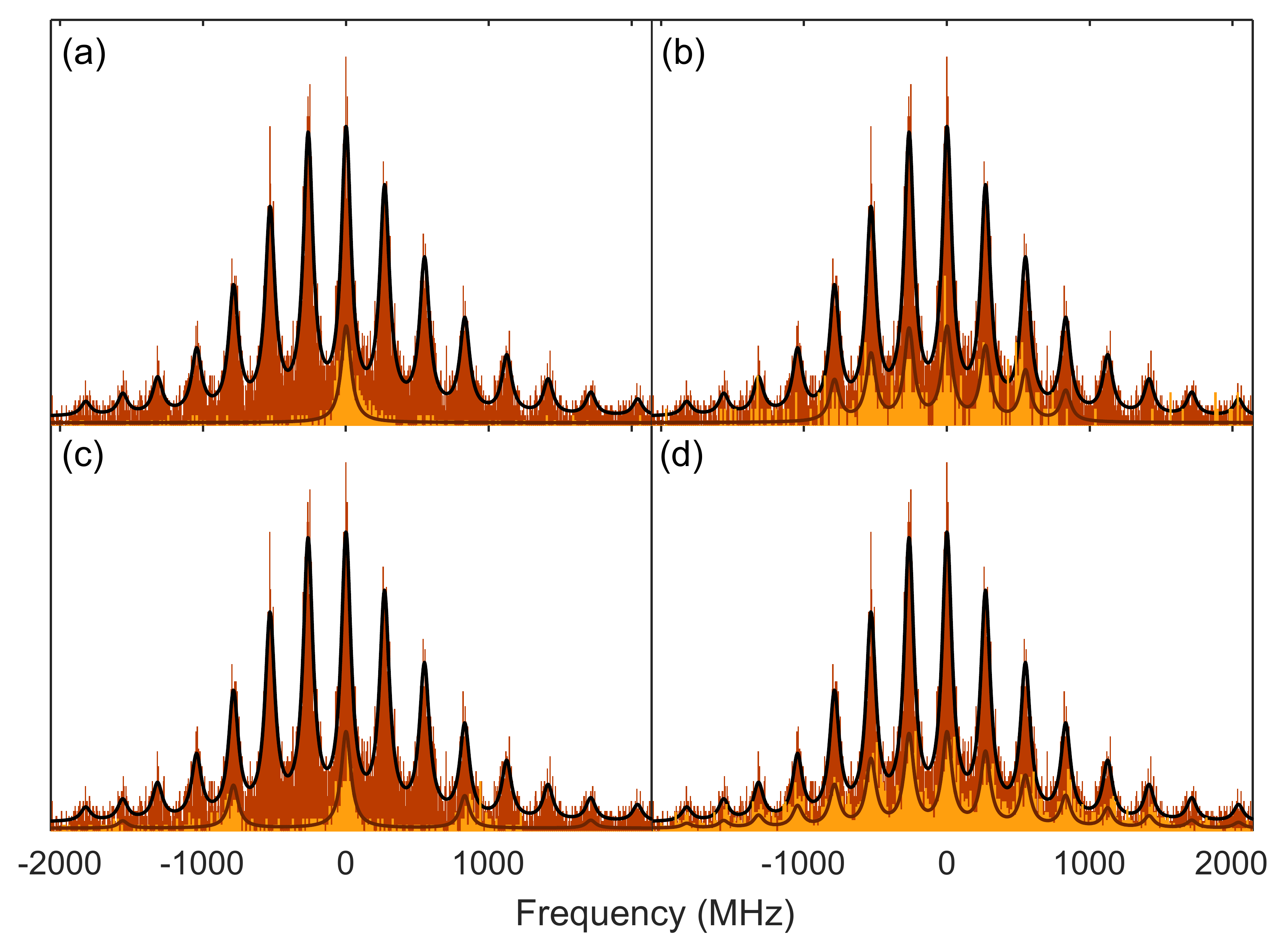}
\caption{Spectrum of the stored photons for different cases: (a) Both EOMs off, single mode case. (b) Just EOM$_1$ on. (c) Just EOM$_2$ on. (d) Both EOM on, fully multimode case. \label{EOMcomb}}
\end{figure}

In the main text we focus our analysis to the case where we store all the modes of the incoming spectrum (Fig. 4(b) of the main text, reported here as Fig. \ref{EOMcomb}(d)) and we compared it with the single mode case (Fig. 4(a) of the main text, reported here as \ref{EOMcomb}(a)). Nevertheless, as shown in Fig. 4(c) and (d) of the main text, we can also switch on just one of the EOMs per measurement. This is shown in Fig \ref{EOMcomb}(b) and \ref{EOMcomb}(c): if we switch on just the EOM$_1$, that we drive with an RF signal at 261.1 MHz, we prepare the AFC for $5\div7$ spectral modes (for the same reason explained in Section \ref{afc_pit_calib}) separated by 261.1 MHz. If we just switch on the EOM$_2$, that we drive at 783.3 MHz, we prepare our AFCs for modes separated by 783.3 MHz.

In order to check it, we repeat the measurements of Fig. 4(a) and (b) of the main text to compare the spectra in the different situations. The results have been fitted according to the criteria introduced in the Section \ref{fits}, keeping the same parameters of the fit (extracted from the brown trace) and suppressing the contribution of the non-stored modes.

\subsection{Scaling of $g^{(2)}_{s,i}(\Delta t)$ with the number of effective modes}
\label{scalingg2}

If we take a look at Fig. 4(d) of the main text we can observe that the value of the cross correlation depends on the number of effective modes (N$_{eM}$). In this section we are going to explain the reason of this behaviour.

The expression for the cross correlation is: $g^{(2)}_{s,i} (\Delta t) = p_{s,i}/(p_s\cdot p_i)$. We call \textit{M} the number of modes generated by our source. For the measurement of Fig. 4(d) of the main text, only a band-pass filter was used in the idler arm ($>$1~nm of linewidth), therefore all the spectrum of the idler photons was measured. N$_{eM}$ is the number of modes addressed in the storage. We call $p_{s,i}$ the coincidence probability for the case of N$_{eM}=1$ and $p_{s(i)}$ the probability to detect independently a signal (idler) photon in a certain time window $\Delta t$, again for N$_{eM}=1$. We can rewrite the expression of the cross correlation as a function of N$_{eM}$ and \textit{M} as: 

\begin{equation}
g^{(2)}_{\textrm{N}_{eM}s,Mi} (\Delta t) = \frac{\textrm{N}_{eM}p_{s,i}}{\textrm{N}_{eM}p_s\cdot Mp_i} = \frac{p_{s,i}}{p_s\cdot Mp_i}
\label{ideal_g2}
\end{equation}

So far, according to the previous expression, the cross correlation should not depend on N$_{eM}$. But we are considering the case of no dark counts in the detectors and no broadband noise. Both of them will add non correlated counts to the measurements. Since the idler detection does not depend on N$_{eM}$, we will consider that it is already intrinsic to the expression $Mp_i$. We can introduce the noise in the signal arm as:
\begin{equation}
g^{(2)}_{\textrm{N}_{eM}s,Mi} (\Delta t) = \frac{\textrm{N}_{eM}p_{s,i}}{(\textrm{N}_{eM}p_s + B) \cdot Mp_i},
\end{equation}
where \textit{B} represents the probability of both detecting dark counts or broadband non-correlated noise. If B is comparable with $p_s$, the effect of this term will only affect our measurements for small values of N$_{eM}$. While our N$_{eM}$ increases we see an asymptotic trend towards the value of eq. \ref{ideal_g2}. This is what we actually observe in Fig. 4(d) of the main text.

\end{document}